\documentclass[fleqn]{llncs}
\usepackage{version}
\usepackage{spis}

\title{Instruction Sequence Notations\\ with Probabilistic Instructions}
\author{J.A. Bergstra \and C.A. Middelburg}
\institute{Informatics Institute, Faculty of Science,
           University of Amsterdam, \\
           Science Park~107, 1098~XG Amsterdam, the Netherlands \\
           \email{J.A.Bergstra@uva.nl,C.A.Middelburg@uva.nl}}

\begin{document}

\maketitle

\begin{abstract}
This paper concerns instruction sequences that contain probabilistic 
instructions, i.e.\ instructions that are themselves probabilistic by 
nature.
We propose several kinds of probabilistic instructions, provide an
informal operational meaning for each of them, and discuss related work.
On purpose, we refrain from providing an ad hoc formal meaning for the
proposed kinds of instructions.
We also discuss the approach of projection semantics, which was 
introduced in earlier work on instruction sequences, in the light of 
probabilistic instruction sequences.
\begin{classcode}
D.1.4, F.1.1, F.1.2.
\end{classcode}
\end{abstract}

\section{Introduction}
\label{sect-intro}

In this paper, we take the first step on a new subject in a line of
research whose working hypothesis is that the notion of an instruction 
sequence is relevant to diverse subjects in computer science (see
e.g.~\cite{BM09k,BM11c,BM08h,BM13a}).
In this line of research, an instruction sequence under execution is
considered to produce a behaviour to be controlled by some execution
environment: each step performed actuates the processing of an
instruction by the execution environment and a reply returned at
completion of the processing determines how the behaviour proceeds.
The term service is used for a component of a system that provides an
execution environment for instruction sequences, and a model of systems
that provide execution environments for instruction sequences is called
an execution architecture.
This paper is concerned with probabilistic instruction sequences.

We use the term probabilistic instruction sequence for an instruction
sequence that contains probabilistic instructions, i.e.\ instructions
that are themselves probabilistic by nature, rather than an instruction
sequence of which the instructions are intended to be processed in a
probabilistic way.
We will propose several kinds of probabilistic instructions, provide
an informal operational meaning for each of them, and discuss related
work.
We will refrain from a formal semantic analysis of the proposed kinds of
probabilistic instructions.
Moreover, we will not claim any form of completeness for the proposed
kinds of probabilistic instructions.
Other convincing kinds might be found in the future.
We will leave unanalysed the topic of probabilistic instruction sequence
processing, which includes all phenomena concerning services and
execution architectures for which probabilistic analysis is necessary.

Viewed from the perspective of machine-execution, execution of a
probabilistic instruction sequence using an execution architecture
without probabilistic features can only be a metaphor.
Execution of a deterministic instruction sequence using an execution
architecture with probabilistic features, i.e.\ an execution
architecture that allows for probabilistic services, is far more
plausible.
Thus, it looks to be that probabilistic instruction sequences find their
true meaning by translation into deterministic instruction sequences for
execution architectures with probabilistic features.
Indeed projection semantics, the approach to define the meaning of
programs which was first presented in~\cite{BL02a}, need not be
compromised when probabilistic instructions are taken into account.

This paper is organized as follows.
First, we set out the scope of the paper (Section~\ref{sect-scope}) and
review the special notation and terminology used in the paper
(Section~\ref{sect-prelim}).
Next, we propose several kinds of probabilistic instructions
(Sections~\ref{sect-basic-test-instrs} and~\ref{sect-jump-instrs}).
Following this, we formulate a thesis on the behaviours produced by 
probabilistic instruction sequences under execution 
(Section~\ref{sect-thesis}) and discuss the approach of projection 
semantics in the light of probabilistic instruction sequences 
(Section~\ref{sect-projectionism}).
We also discuss related work (Section~\ref{sect-related-work}) and
make some concluding remarks (Section~\ref{sect-concl}).

In the current version of this paper, we moreover mention some outcomes 
of a sequel to the work reported upon in this paper
(Section~\ref{sect-sequel}).

\section{On the Scope of this Paper}
\label{sect-scope}

We go into the scope of the paper and clarify its restrictions by giving
the original motives.
However, the first version of this paper triggered off work on the 
behaviour of probabilistic instruction sequences under execution and 
outcomes of that work invalidated some of the arguments used to motivate 
the restrictions.
The relevant outcomes of the work concerned will be mentioned in 
Section~\ref{sect-sequel}.

We will propose several kinds of probabilistic instructions, chosen
because of their superficial similarity with kinds of deterministic
instructions known from \PGA\ (ProGram Algebra)~\cite{BL02a}, 
\PGLD\ (ProGramming Language D) with indirect jumps~\cite{BM07e}, 
C (Code)~\cite{BP09a}, and other similar notations and not because any 
computational intuition about them is known or assumed.
For each of these kinds, we will provide an informal operational
meaning.
Moreover, we will exemplify the possibility that the proposed unbounded 
probabilistic jump instructions are simulated by means of bounded 
probabilistic test instructions and bounded deterministic jump 
instructions.
We will also refer to related work that introduces something similar to
what we call a probabilistic instruction and connect the proposed kinds
of probabilistic instructions with similar features found in related
work.

We will refrain from a formal semantic analysis of the proposed kinds of
probabilistic instructions.
When we started with the work reported upon in this paper, the reasons 
for doing so were as follows:
\begin{itemize}
\item
In the non-probabilistic case, the subject reduces to the semantics of
\PGA.
Although it seems obvious at first sight, different models, reflecting
different levels of abstraction, can and have been distinguished (see
e.g.~\cite{BL02a}).
Probabilities introduce a further ramification.
\item
What we consider sensible is to analyse this double ramification fully.
What we consider less useful is to provide one specific collection of
design decisions and working out its details as a proof of concept.
\item
We notice that for process algebra the ramification of semantic options
after the incorporation of probabilistic features is remarkable, and
even frustrating (see e.g.~\cite{GSS95a,JLY01a}).
There is no reason to expect that the situation is much simpler here.
\item
Once that a semantic strategy is mainly judged on its preparedness for a
setting with multi-threading, the subject becomes intrinsically complex
-- like the preparedness for a setting with arbitrary interleaving
complicates the semantic modeling of deterministic processes in process
algebra.
\item
We believe that a choice for a catalogue of kinds of probabilistic
instructions can be made beforehand.
Even if that choice will turn out to be wrong, because prolonged
forthcoming semantic analysis may give rise to new, more natural, kinds
of probabilistic instructions, it can at this stage best be driven by
direct intuitions.
\end{itemize}

In this paper, we will leave unanalysed the topic of probabilistic 
instruction sequence processing, i.e.\ probabilistic processing of 
instruction sequences, which includes all phenomena concerning services 
and concerning execution architectures for which probabilistic analysis 
is necessary.
At the same time, we admit that probabilistic instruction sequence
processing is a much more substantial topic than probabilistic
instruction sequences, because of its machine-oriented scope.
We take the line that a probabilistic instruction sequence finds its
operational meaning by translation into a deterministic instruction
sequence and execution using an execution architecture with
probabilistic features.

\section{Preliminaries}
\label{sect-prelim}

In the remainder of this paper, we will use the notation and terminology
regarding instructions and instruction sequences from \PGA.
The mathematical structure that we will use for quantities is a signed
cancellation meadow.
That is why we briefly review \PGA\ and signed cancellation meadows in
this section.

In \PGA, it is assumed that a fixed but arbitrary set of
\emph{basic instructions} has been given.
The \emph{primitive instructions} of \PGA\ are the basic instructions
and in addition:
\begin{itemize}
\item
for each basic instruction $a$,
a \emph{positive test instruction} $\ptst{a}$;
\item
for each basic instruction $a$,
a \emph{negative test instruction} $\ntst{a}$;
\item
for each natural number $l$,
a \emph{forward jump instruction} $\fjmp{l}$;
\item
a \emph{termination instruction} $\halt$.
\end{itemize}

The intuition is that the execution of a basic instruction $a$ produces 
either $\True$ or $\False$ at its completion.
In the case of a positive test instruction $\ptst{a}$, $a$ is executed
and execution proceeds with the next primitive instruction if $\True$ is
produced.
Other\-wise, the next primitive instruction is skipped and execution
proceeds with the primitive instruction following the skipped one.
If there is no next instruction to be executed, execution becomes 
inactive.
In the case of a negative test instruction $\ntst{a}$, the role of the
value produced is reversed.
In the case of a plain basic instruction $a$, execution always proceeds
as if $\True$ is produced.
The effect of a forward jump instruction $\fjmp{l}$ is that execution
proceeds with the $l$-th next instruction.
If $l$ equals $0$ or the $l$-th next instruction does not exist,
execution becomes inactive.
The effect of the termination instruction $\halt$ is that execution
terminates.

The constants of \PGA\ are the primitive instructions and the operators
of \PGA\ are:
\begin{itemize}
\item
the binary \emph{concatenation} operator $\conc$\,;
\item
the unary \emph{repetition} operator $\rep$\,.
\end{itemize}
Terms are built as usual.
We use infix notation for the concatenation operator and postfix
notation for the repetition operator.

A closed \PGA\ term is considered to denote a non-empty, finite or
periodic infinite sequence of primitive instructions.%
\footnote
{A periodic infinite sequence is an infinite sequence with only finitely
 many distinct suffixes.}
Closed \PGA\ terms are considered equal if they denote the same
instruction sequence.
The axioms for instruction sequence equivalence are given
in~\cite{BL02a}.
The \emph{unfolding} equation $X\rep = X \conc X\rep$ is derivable from
those equations.
Moreover, each closed \PGA\ term is derivably equal to one of the form
$P$ or $P \conc Q\rep$, where $P$ and $Q$ are closed \PGA\ terms in
which the repetition operator does not occur.

In~\cite{BL02a}, \PGA\ is extended with a \emph{unit instruction}
operator $\uiop$ which turns sequences of instructions into single
instructions.
The result is called \PGA$_\mathrm{u}$.
In~\cite{Pon02a}, the meaning of \PGA$_\mathrm{u}$ programs is described
by a translation from \PGA$_\mathrm{u}$ programs into \PGA\ programs.

In the sequel, the following additional assumption is made: a fixed but
arbitrary set of \emph{foci} and a fixed but arbitrary set of
\emph{methods} have been given.
Moreover, we will use $f.m$, where $f$ is a focus and $m$ is a method,
as a general notation for basic instructions.
In $f.m$, $m$ is the instruction proper and $f$ is the name of the
service that is designated to process $m$.

The signature of signed cancellation meadows consists of the following
constants and operators:
\begin{itemize}
\item
the constants $0$ and $1$;
\item
the binary \emph{addition} operator ${} +$ {};
\item
the binary \emph{multiplication} operator ${} \mul {}$;
\item
the unary \emph{additive inverse} operator $- {}$;
\item
the unary \emph{multiplicative inverse} operator ${}\minv$;
\item
the unary \emph{signum} operator $\sign$.
\end{itemize}

Terms are build as usual.
We use infix notation for the binary operators ${} + {}$ and
${} \mul {}$, prefix notation for the unary operator $- {}$, and postfix
notation for the unary operator ${}\minv$.
We use the usual precedence convention to reduce the need for
parentheses.
We introduce subtraction and division as abbreviations:
$p - q$ abbreviates $p + (-q)$ and
$p / q$ abbreviates $p \mul (q\minv)$.
We use the notation $\num{n}$ for numerals and the notation $p^n$ for
exponentiation with a natural number as exponent.
The term $\num{n}$ is inductively defined as follows: $\num{0} = 0$ and
$\num{n+1} = \num{n} + 1$.
The term $p^n$ is inductively defined as follows: $p^0 = 1$ and
$p^{n+1} = p^n \mul p$.

The constants and operators from the signature of signed cancellation
meadows are adopted from rational arithmetic, which gives an appropriate
intuition about these constants and operators.
The equational theories of signed cancellation meadows is given
in~\cite{BBP13a}.
In signed cancellation meadows, the functions $\min$ and $\max$ have a
simple definition (see also~\cite{BBP13a}).

A signed cancellation meadow is a cancellation meadow expanded with a
signum operation.
The prime example of cancellation meadows is the field of rational
numbers with the multiplicative inverse operation made total by imposing
that the multiplicative inverse of zero is zero, see e.g.~\cite{BT07a}.

\section{Probabilistic Basic and Test Instructions}
\label{sect-basic-test-instrs}

In this section, we propose several kinds of probabilistic basic and
test instructions.
It is assumed that a fixed but arbitrary signed cancellation meadow
$\fM$ has been given.
Moreover, we write $\mkprob{q}$, where $q \in \fM$, for
$\max(0,\min(1,q))$.

We propose the following \emph{probabilistic basic instructions}:
\begin{itemize}
\item
$\prbsc{}$, which  produces $\True$ with probability $1/2$ and $\False$
with probability $1/2$;
\item
$\prbsc{q}$, which produces $\True$ with probability $\mkprob{q}$ and 
$\False$ with probability $1 - \mkprob{q}$,\linebreak[2] 
for $q \in \fM$.
\end{itemize}
The probabilistic basic instructions have no side-effect on a state.

The basic instruction $\prbsc{}$ can be looked upon as a shorthand for
$\prbsc{1/2}$.
We distinguish between $\prbsc{}$ and $\prbsc{1/2}$ for reason of
putting the emphasis on the fact that it is not necessary to bring in a
notation for quantities ranging from $0$ to $1$ in order to design
probabilistic instructions.

Once that probabilistic basic instructions of the form $\prbsc{q}$ are
chosen, an unbounded ramification of options for the notation of
quantities is opened up.
We will assume that closed terms over the signature of signed
cancellation meadows are used to denote quantities.
Instructions such as $\prbsc{\sqrt{1+1}}$ are implicit in the form
$\prbsc{q}$, assuming that it is known how to view $\sqrt{\,}$ as a
notational extension of signed cancellation meadows (see
e.g.~\cite{BBP13a}).

Like all basic instructions, each probabilistic basic instruction give
rise to two \emph{probabilistic test instructions}:
\begin{itemize}
\item
$\prbsc{}$ gives rise to $\ptst{\prbsc{}}$ and $\ntst{\prbsc{}}$;
\item
$\prbsc{q}$ gives rise to $\ptst{\prbsc{q}}$ and $\ntst{\prbsc{q}}$.
\end{itemize}
Probabilistic primitive instructions of the form $\ptst{\prbsc{q}}$ and
$\ntst{\prbsc{q}}$ can be considered probabilistic branch instructions
where $q$ is the probability that the branch is not taken and taken,
respectively, and likewise the probabilistic primitive instructions
$\ptst{\prbsc{}}$ and $\ntst{\prbsc{}}$.

We find that the primitive instructions $\prbsc{}$ and $\prbsc{q}$ can
be replaced by $\fjmp{1}$ without loss of (intuitive) meaning.
Of course, in a resource aware model, $\fjmp{1}$ may be much cheaper
than $\prbsc{q}$, especially if $q$ is hard to compute.
Suppose that $\prbsc{q}$ is realized at a lower level by means of
$\prbsc{}$, which is possible, and suppose that $q$ is a computable real
number.
The question arises whether the expectation of the time to execute
$\prbsc{q}$ is finite.

To exemplify the possibility that $\prbsc{q}$ is realized by means of
$\prbsc{}$ in the case where $q$ is a rational number, we look at the
following probabilistic instruction sequences:
\begin{ldispl}
\ntst{\prbsc{2/3}} \conc \fjmp{3} \conc
a \conc \halt \conc b \conc \halt\;,
\\
(\ptst{\prbsc{}} \conc \fjmp{3} \conc a \conc \halt \conc
 \ptst{\prbsc{}} \conc \fjmp{3} \conc b \conc \halt)\rep\;.
\end{ldispl}
It is easy to see that these instruction sequences produce on execution
the same behaviour: with probability $2/3$, first $a$ is performed and
then termination follows; and with probability $1/3$, first $b$ is
performed and then termination follows.
In the case of computable real numbers other than rational numbers, use
must be made of a service that does duty for the tape of a Turing
machine (such a service is, for example, described in~\cite{BP04a}).

Let $q \in \fM$, and let $\random(q)$ be a service with a method $\get$
whose reply is $\True$ with probability $\mkprob{q}$ and $\False$ with 
probability $1 - \mkprob{q}$.
Then a reasonable view on the meaning of the probabilistic primitive
instructions $\prbsc{q}$, $\ptst{\prbsc{q}}$ and $\ntst{\prbsc{q}}$ is
that they are translated into the deterministic primitive instructions
$\random(q).\get$, $\ptst{\random(q).\get}$ and
$\ntst{\random(q).\get}$, respectively, and executed using an execution
architecture that provides the probabilistic service $\random(q)$.
Another option is possible here: instead of a different service
$\random(q)$ for each $q \in [0,1]$ and a single method $\get$, we could
have a single service $\random$ with a different method $\get(q)$ for
each $q \in [0,1]$.
In the latter case, $\prbsc{q}$, $\ptst{\prbsc{q}}$ and
$\ntst{\prbsc{q}}$ would be translated into the deterministic primitive
instructions $\random.\get(q)$, $\ptst{\random.\get(q)}$ and
$\ntst{\random.\get(q)}$.

\section{Probabilistic Jump Instructions}
\label{sect-jump-instrs}

In this section, we propose several kinds of probabilistic jump
instructions.
It is assumed that the signed cancellation meadow $\fM$ has been
expanded with an operation $\Nat$ such that, for all $q \in \fM$,
$\Nat(q) = 0$ iff $q = \num{n}$ for some $n \in \Nat$.\linebreak[2]
We write $\numinv{l}$, where $l \in \fM$ is such that $\Nat(l) = 0$, for
the unique $n \in \Nat$ such that $l = \num{n}$.

We propose the following \emph{probabilistic jump instructions}:
\begin{itemize}
\item
$\prjmpu{k}$, having the same effect as $\fjmp{j}$ with probability
$1/k$ for $j \in [1,\numinv{k}]$,\linebreak[2] for $k \in \fM$ with
$\Nat(k) = 0$;
\item
$\prjmpg{q}{k}$, having the same effect as $\fjmp{j}$ with probability
$\mkprob{q} \mul (1 - \mkprob{q})^{j-1}$ for $j \in [1,\numinv{k}]$, for
$q \in \fM$ and $k \in \fM$ with $\Nat(k) = 0$;
\item
$\prjmpgu{q}{l}$, having the same effect as $\fjmp{\numinv{l} \mul j}$
with probability $\mkprob{q} \mul (1 - \mkprob{q})^{j-1}$ for
$j \in [1,\infty)$, for $q \in \fM$ and $l \in \fM$ with $\Nat(l) = 0$.
\end{itemize}
The letter U in $\prjmpu{k}$ indicates a uniform probability 
distribution, and the letter G in $\prjmpg{q}{k}$ and $\prjmpgu{q}{l}$
indicates a geometric probability distribution.
Instructions of the forms $\prjmpu{k}$ and $\prjmpg{q}{k}$ are bounded
probabilistic jump instructions, whereas instructions of the form
$\prjmpgu{q}{l}$ are unbounded probabilistic jump instructions.

Like in the case of the probabilistic basic instructions, we propose
in addition the following probabilistic jump instructions:
\begin{itemize}
\item
$\prjmpg{}{k}$ as the special case of $\prjmpg{q}{k}$ where $q = 1/2$;
\item
$\prjmpgu{}{l}$ as the special case of $\prjmpgu{q}{l}$ where $q = 1/2$.
\end{itemize}

We believe that it must be possible to eliminate all probabilistic jump 
instructions.
In particular, we believe that it must be possible to eliminate all
unbounded probabilistic jump instructions.
This belief can be understood as the judgement that it is reasonable to
expect from a semantic model of probabilistic instruction sequences that
the following identity and similar ones hold:
\begin{ldispl}
\ptst{a} \conc \prjmpgu{}{2} \conc (\ptst{b} \conc \halt \conc c)\rep
 = {}
\\
\ptst{a} \conc \ptst{\prbsc{}} \conc \fjmp{8} \conc \fjmp{10} \conc {}
\\ \quad
(\ptst{b} \conc \fjmp{5} \conc \fjmp{10} \conc \ptst{\prbsc{}} \conc
 \fjmp{8} \conc \fjmp{10} \conc {}
\\ \quad \phantom{({+}}\hsp{0.125}
 \halt \conc \fjmp{5} \conc \fjmp{10} \conc \ptst{\prbsc{}} \conc
 \fjmp{8} \conc \fjmp{10} \conc {}
\\ \quad \phantom{({+}}
 c \conc \fjmp{5} \conc \fjmp{10} \conc \ptst{\prbsc{}} \conc
 \fjmp{8} \conc \fjmp{10}
)\rep\;.
\end{ldispl}
Taking this identity and similar ones as our point of departure, the
question arises what is the most simple model that justifies them.
A more general question is whether instruction sequences with unbounded
probabilistic jump instructions can be translated into ones without
probabilistic jump instructions provided it does not bother us that the
instruction sequences may become much longer (e.g.\ expectation of the
length bounded, but worst case length unbounded).

\section{The Probabilistic Process Algebra Thesis}
\label{sect-thesis}

In the absence of probabilistic instructions, threads as considered in
\BTA\ (Basic\linebreak[2] Thread Algebra)~\cite{BL02a} or its extension 
with thread-service interaction~\cite{BP02a} can be used to model the 
behaviours produced by instruction sequences under execution.%
\footnote
{In~\cite{BL02a}, \BTA\ is introduced under the name \BPPA\ 
 (Basic Polarized Process Algebra).
}
Processes as considered in general process algebras such as
\ACP~\cite{BW90}, CCS~\cite{Mil89} and CSP~\cite{Hoa85} can be used as
well, but they give rise to a more complicated modeling of the
behaviours of instruction sequences under execution (see
e.g.~\cite{BM11c}).

In the presence of probabilistic instructions, we would need a
probabilistic thread algebra, i.e.\ a variant of \BTA\ or its extension 
with thread-service interaction that covers probabilistic behaviours.
When we started with the work reported upon in this paper, it appeared 
that any probabilistic thread algebra is inherently more complicated to 
such an extent that the advantage of not using a general process algebra 
evaporates.
Moreover, it appeared that any probabilistic thread algebra requires
justification by means of an appropriate probabilistic process algebra.
This led us to the following thesis:
\begin{thesis}
Modeling the behaviours produced by probabilistic instruction sequences
under execution is a matter of using directly processes as considered in
some probabilistic process algebra.
\end{thesis}

A probabilistic thread algebra has to cover the interaction between
instruction sequence behaviours and services.
Two mechanisms are involved in that.
They are called the use mechanism and the apply mechanism (see
e.g.~\cite{BP02a}).
The difference between them is a matter of perspective: the former is
concerned with the effect of services on behaviours of instruction
sequences and therefore produces behaviours, whereas the latter is
concerned with the effect of instruction sequence behaviours on services
and therefore produces services.
When we started with the work reported upon in this paper, it appeared 
that the use mechanism would make the development of a probabilistic 
thread algebra very intricate.

The first version of this paper triggered off work on the behaviour of 
probabilistic instruction sequences under execution by which the thesis
stated above is refuted. 
The ultimate point is that meanwhile an appropriate and relatively 
simple probabilistic thread algebra has been devised (see~\cite{BM14d}).
Moreover, our original expectations about probabilistic process algebras
turned out to be too high.

The first probabilistic process algebra is presented in~\cite{GJS90a} 
and the first probabilistic process algebra with an asynchronous 
parallel composition operator is presented in~\cite{BBS95a}.
A recent overview of the work on probabilistic process algebras after 
that is given in~\cite{LN04a}.
This overview shows that the multitude of semantic ideas applied and the 
multitude of variants of certain operators devised have kept growing,
and that convergence is far away.
In other words, there is little well-established yet.
In particular, for modeling the behaviours produced by probabilistic 
instruction sequences, we need operators for probabilistic choice, 
asynchronous parallel composition, and abstraction from internal 
actions. 
For this case, the attempts by one research group during about ten years 
to develop a satisfactory ACP-like process algebra (see 
e.g.~\cite{AB01a,ABW06a,AG09a}) have finally led to a promising process 
algebra. 
However, it is not yet clear whether the process algebra concerned will 
become well-established.

All this means that a justification of the above-mentioned probabilistic 
thread algebra by means of an appropriate probabilistic process algebra 
will be of a provisional nature for the time being.

\section{Discussion of Projectionism}
\label{sect-projectionism}

Notice that once we move from deterministic instructions to
probabilistic instructions, instruction sequence becomes an
indispensable concept.
Instruction sequences cannot be replaced by threads or processes without
taking potentially premature design decisions.
In preceding sections, however, we have outlined how instruction 
sequences with the different kinds of probabilistic instructions can be 
translated into instruction sequences without them.
Therefore, it is a reasonable to claim that, like for deterministic
instruction sequence notations, all probabilistic instruction sequence
notations can be provided with a probabilistic semantics by translation
of the instruction sequences concerned into appropriate single-pass
instruction sequences.
Thus, we have made it plausible that projectionism is feasible for
probabilistic instruction sequences.

\emph{Projectionism} is the point of view that:
\begin{itemize}
\item
any instruction sequence $P$, and more general even any program $P$, first
and for all represents a single-pass instruction sequence as considered
in \PGA;
\item
this single-pass instruction sequence, found by a translation called a
projection, represents in a natural and preferred way what is supposed
to take place on execution of $P$;
\item
\PGA\ provides the preferred notation for single-pass instruction
sequences.
\end{itemize}
In a rigid form, as in~\cite{BL02a}, projectionism provides a definition
of what constitutes a program.

The fact that projectionism is feasible for probabilistic instruction
sequences, does not imply that it is uncomplicated.
To give an idea of the complications that may arise, we will sketch below
found challenges for projectionism.

First, we introduce some special notation.
Let \PGN\ be a program notation.
Then we write $\npga{\PGN}$ for the projection function that gives, for
each program $P$ in~\PGN, the closed \PGA\ terms that denotes the
single-pass instruction sequence that produces on execution the same
behaviour as $P$.

We have found the following challenges for projectionism:
\begin{itemize}
\item
\emph{Explosion of size}.\,\,
If $\npga{\PGN}(P)$ is much longer than $P$, then the requirement that
it represents in a natural way what is supposed to take place on
execution of $P$ is challenged.
For example, if the primitive instructions of \PGN\ include instructions
to set and test up to $n$ Boolean registers, then the projection to
$\npga{\PGN}(P)$ may give rise to a combinatorial explosion of size.
In such cases, the usual compromise is to permit single-pass instruction
sequences to make use of services (see e.g.~\cite{BP02a}).
\item
\emph{Degradation of performance}.\,\,
If $\npga{\PGN}(P)$'s natural execution is much slower than $P$'s
execution, supposing a clear operational understanding of $P$, then the
requirement that it represents in a natural way what is supposed to take
place on execution of $P$ is challenged.
For example, if the primitive instructions of \PGN\ include indirect
jump instructions, then the projection to $\npga{\PGN}(P)$ may give rise
to a degradation of performance (see e.g.~\cite{BM07e}).
\item
\emph{Incompatibility of services}.\,\,
If $\npga{\PGN}(P)$ has to make use of services that are not
deterministic, then the requirement that it represents in a natural way
what is supposed to take place on execution of $P$ is challenged.
For example, if the primitive instructions of \PGN\ include instructions
of the form $\ptst{\prbsc{q}}$ or $\ntst{\prbsc{q}}$, then $P$ cannot be
projected to a single-pass instruction sequence without the use of
probabilistic services.
In this case, either probabilistic services must be permitted or
probabilistic instruction sequences must not be considered programs.
\item
\emph{Complexity of projection description}.\,\,
The description of $\npga{\PGN}$ may be so complex that it defeats
$\npga{\PGN}(P)$'s purpose of being a natural explanation of what is
supposed to take place on execution of $P$.
For example, the projection semantics given for recursion
in~\cite{BB06a} suffers from this kind of complexity when compared with
the conventional denotational semantics.
In such cases, projectionism may be maintained conceptually, but
rejected pragmatically.
\item
\emph{Aesthetic degradation}.\,\,
In $\npga{\PGN}(P)$, something elegant may have been replaced by nasty
details.
For example, if \PGN\ provides guarded commands, then $\npga{\PGN}(P)$,
which will be much more detailed, might be considered to exhibit signs
of aesthetic degradation.
This challenge is probably the most serious one, provided we accept that
such elegant features belong to program notations.
Of course, it may be decided to ignore aesthetic criteria altogether.
However, more often than not, they have both conceptual and pragmatic
importance.
\end{itemize}

One might be of the opinion that conceptual projectionism can accept
explosion of size and/or degradation of performance.
We do not share this opinion: both challenges require a more drastic
response than a mere shift from a pragmatic to a conceptual
understanding of projectionism.
This drastic response may include viewing certain mechanisms as
intrinsically indispensable for either execution performance or program
compactness.
For example, it is reasonable to consider the basic instructions of the
form $\prbsc{q}$, where $q$ is a computable real number, indispensable
if the expectations of the times to execute their realizations by means
of $\prbsc{}$ are not all finite.

Nevertheless, projectionism looks to be reasonable for probabilistic
programs: they can be projected adequately to deterministic single-pass
instruction sequences for an execution architecture with probabilistic
services.

\section{Related Work}
\label{sect-related-work}

In~\cite{SPH84a}, a notation for probabilistic programs is introduced in
which we can write, for example,
$\mathrm{random}(p \mul \delta_0 + q \mul \delta_1)$.
In general, $\mathrm{random}(\lambda)$ produces a value accord\-ing to
the probability distribution $\lambda$.
In this case, $\delta_i$ is the probability distribution that gives
probability $1$ to $i$ and probability $0$ to other values.
Thus, for $p + q = 1$, $p \mul \delta_0 + q \mul \delta_1$ is the
probability distribution that gives probability $p$ to $0$, probability
$q$ to $1$, and probability $0$ to other values.
Clearly, $\mathrm{random}(p \mul \delta_0 + q \mul \delta_1)$
corresponds to $\prbsc{p}$.
Moreover, using this kind of notation, we could write
$\fjmp{(\frac{1}{k} \mul (\delta_1 + \cdots + \delta_{\numinv{k}}))}$
for $\prjmpu{k}$ and
$\fjmp{(\mkprob{q} \mul \delta_1 +
        \mkprob{q} \mul (1 - \mkprob{q}) \mul \delta_2 + \cdots +
        \mkprob{q} \mul (1 - \mkprob{q})^{k - 1} \mul \delta_k)}$
for $\prjmpg{q}{k}$.

In much work on probabilistic programming, see
e.g.~\cite{HSM97a,MM01a,MMS96a}, we find the binary probabilistic choice
operator $\prchc{p}$ (for $p \in [0,1]$).
This operator chooses between its operands, taking its left operand with
probability $p$.
Clearly, $P \prchc{p} Q$ can be taken as abbreviations for
$\ptst{\prbsc{p}} \conc \uiop(P \conc \fjmp{2}) \conc \uiop(Q)$.
This kind of primitives dates back to~\cite{Koz85a} at least.

Quite related, but from a different perspective, is the $\toss$
primitive introduced in~\cite{CCMS07a}.
The intuition is that $\toss(\nm{bm},p)$ assigns to the Boolean memory
cell $\nm{bm}$ the value $\True$ with probability $\mkprob{p}$ and the
value $\False$ with probability $1 - \mkprob{p}$.
This means that $\toss(\nm{bm},p)$ has a side-effect on a state, which
we understand as making use of a service.
In other words, $\toss(\nm{bm},p)$ corresponds to a deterministic
instruction intended to be processed by a probabilistic service.

Common in probabilistic programming are assignments of values randomly
chosen from some interval of natural numbers to program variables (see
e.g.~\cite{Sch02a}).
Clearly, such random assignments correspond also to deterministic
instructions intended to be processed by probabilistic services.
Suppose that $x{=}i$ is a primitive instruction for assigning value $i$
to program variable $x$.
Then we can write:
$\prjmpu{k} \conc \uiop(x{=}1 \conc \fjmp{k}) \conc
                  \uiop(x{=}2 \conc \fjmp{k{-}1}) \conc \ldots \conc
                  \uiop(x{=}k \conc \fjmp{1})$.
This is a realistic representation of the assignment to $x$ of a value
randomly chosen from $\set{1,\ldots,k}$.
However, it is clear that this way of representing random assignments
leads to an exponential blow up in the size of any concrete instruction
sequence representation, provided the concrete representation of $k$ is
its decimal representation.

The refinement oriented theory of programs uses demonic choice,
usually written $\sqcap$, as a primitive (see e.g.~\cite{MM01a,MS08a}).
A demonic choice can be regarded as a probabilistic choice with unknown
probabilities.
Demonic choice could be written $\ptst{{\sqcap}}$ in a \PGA-like
notation.
However, a primitive instruction corresponding to demonic choice is not
reasonable: no mechanism for the execution of $\ptst{{\sqcap}}$ is
conceivable.
Demonic choice exists in the world of specifications, but not in the
world of instruction sequences.
This is definitely different with $\ptst{\prbsc{p}}$, because a
mechanism for its execution is conceivable.

Features similar to probabilistic jump instructions are not common in
probabilistic programming.
To our knowledge, the mention of probabilistic goto statements of the
form $\prgoto\; \set{l_1,l_2}$ in~\cite{APZ03a} is the only mention of
a similar feature in the computer science literature.
The intuition is that $\prgoto\; \set{l_1,l_2}$, where $l_1$ and $l_2$
are labels, has the same effect as $\goto\; l_1$ with probability $1/2$ 
and has the same effect as $\goto\; l_2$ with probability $1/2$.
Clearly, this corresponds to a probabilistic jump
instruction of the form $\prjmpu{1/2}$.

It appears that quantum computing has something to offer that cannot be
obtained by conventional computing: it makes a stateless generator of
random bits available (see e.g.~\cite{Gay06a,PJ06a}).
By that quantum computing indeed provides a justification of
$\ptst{\prbsc{1/2}}$ as a probabilistic instruction.

\section{On the Sequel to this Paper}
\label{sect-sequel}

The first version of this paper triggered off work on the behaviour of 
probabilistic instruction sequences under execution and outcomes of that 
work invalidated some of the arguments used to motivate the restricted 
scope of this paper.
In this section, we mention the relevant outcomes of that work.

After the first version of this paper (i.e.~\cite{BM09f}) appeared, it
was found that:
\begin{itemize}
\item
The different levels of abstraction that have been distinguished in the
non-probabilistic case can be distinguished in the probabilistic case 
as well and only the models at the level of the behaviour of instruction 
sequences under execution are not essentially the same.
\item
The semantic options at the behavioural level after the incorporation of 
probabilistic features are limited because of 
(a)~the orientation towards behaviours of a special kind and
(b)~the semantic constraints induced by the informal explanations of
the enumerated kinds of probabilistic instructions and the desired
elimination property of all but one kind.
\item
For the same reasons, preparedness for a setting with multi-threading 
does not really complicate matters. 
\end{itemize}
The current state of affairs is as follows:
\begin{itemize}
\item
\BTA\ and its extension with thread-service interaction, which are used 
to describe the behaviour of instruction sequences under execution in 
the non-probabilistic case, have been extended with probabilistic 
features in~\cite{BM14d}.
\item
In~\cite{BM14d}, we have added the probabilistic basic and test 
instructions proposed in Section~\ref{sect-basic-test-instrs} to PGLB
(ProGramming Language B), an instruction sequence notation rooted in 
\PGA\ and close to existing assembly languages, and have given a formal 
definition of the behaviours produced by the instruction sequences from 
the resulting instruction sequence notation in terms of non-probabilistic 
instructions and probabilistic services.
\item
The bounded probabilistic jump instructions proposed in 
Section~\ref{sect-jump-instrs} can be given a behavioural semantics in 
the same vein.
The unbounded probabilistic jump instructions fail to have a natural 
behavioural semantics in the setting of \PGA\ because infinite 
instruction sequences are restricted to eventually periodic ones.
\item
The extension of \BTA\ with multi-threading, has been generalized to 
probabilistic multi-threading in~\cite{BM14d} as well.
\end{itemize}

\section{Conclusions}
\label{sect-concl}

We have made a notational proposal of probabilistic instructions with an
informal semantics.
By that we have contrasted probabilistic instructions in an execution
architecture with deterministic services with deterministic instructions
in an execution architecture with partly probabilistic services.
The history of the proposed kinds of instructions can be traced.

We have refrained from an ad hoc formal semantic analysis of the
proposed kinds of instructions.
There are many solid semantic options, so many and so complex that
another more distant analysis is necessary in advance to create a clear
framework for the semantic analysis in question.

The grounds of this work are our conceptions of what a theory of
probabilistic instruction sequences and a complementary theory of
probabilistic instruction sequence processing (i.e.\ execution
architectures with probabilistic services) will lead to:
\begin{itemize}
\item
comprehensible explanations of relevant probabilistic algorithms, such
as the Miller-Rabin probabilistic primality test~\cite{Rab76a}, with
precise descriptions of the kinds of instructions and services involved
in them;
\item
a solid account of pseudo-random Boolean values and pseudo-random
numbers;
\item
a thorough exposition of the different semantic options for
probabilistic instruction sequences;
\item
explanations of relevant quantum algorithms, such as Shor's integer
factorization algorithm~\cite{Sho94a}, by first giving a clarifying
analysis in terms of probabilistic instruction sequences or execution
architectures with probabilistic services and only then showing how
certain services in principle can be realized very efficiently with
quantum computing.
\end{itemize}

Projectionism looks to be reasonable for probabilistic programs: they
can be projected adequately to deterministic single-pass instruction
sequences for an execution architecture with appropriate probabilistic
services.
At present, it is not entirely clear whether this extends to quantum
programs.

\bibliographystyle{splncs03}
\bibliography{IS}

\end{document}